\documentclass[11pt]{article}
\usepackage{amssymb}
\usepackage{txfonts}%
\usepackage{graphicx}
\usepackage{epsfig}
\usepackage{graphics,color}
\usepackage{amsbsy}

\newcommand{\be}{\begin{equation}}
 \newcommand{\ee}{\end{equation}}
 \newcommand{\bse}{\begin{subequations}}
 \newcommand{\ese}{\end{subequations}}
 \newcommand{\bea}{\begin{eqnarray}}
 \newcommand{\eea}{\end{eqnarray}}

\newcommand{\bean}{\begin{eqnarray*}}
\newcommand{\eean}{\end{eqnarray*}}

\begin{document}

\begin{center}

 {\bf Particle production in relativistic $pp$($\overline{p}$) and $AA$
collisions at RHIC and LHC energies with Tsallis statistics using
the two-cylindrical multisource thermal model} \vskip1.0cm

Bao-Chun Li\footnote{libc2010@163.com, s6109@sxu.edu.cn}, Ya-Zhou
Wang, Fu-Hu Liu, Xin-Jian Wen and You-Er Dong

{\small {\it Department of Physics, Shanxi University, Taiyuan,
Shanxi 030006, China}}
\end{center}

\begin{abstract}

An improved Tsallis statistics is implemented  in a multisource
thermal model to describe systematically pseudorapidity spectra of
charged particles produced in relativistic nucleon-nucleon ($pp$ or
$p\overline{p}$) collisions at various collision energies and in
relativistic nucleus-nucleus ($AA$) collisions at different energies
with different centralities. The results with Tsallis statistics
using the two-cylindrical multisource thermal model are in good
agreement with the experimental data measured at RHIC and LHC
energies. It is found that the rapidity shifts of longitudinal
sources increase linearly with collision energies  and centralities
in the framework. According to the laws, we also give a prediction
of the pseudorapidity distributions in $pp$($\overline{p}$)
collisions at higher
energies.\\
\\
PACS number(s): 25.75.-q, 25.75.Dw, 24.10.Pa

\end{abstract}

\vskip1.0cm

\newpage

{\section{INTRODUCTION}}

Multiparticle production is an important experimental phenomenon at
the Relativistic Heavy Ion Collider (RHIC) in Brookhaven National
Laboratory.  In Au+Au collisions, final-state particle yields per
unity of rapidity integrated over transverse momentum $p_T$ ranges
have provided the information of the temperature $T$ and chemical
potential $\mu$ at the chemical freeze-out by using a statistical
investigation~\cite{Adams:2003xp}. It brings valuable insight into
 properties of quark-gluon plasma (QGP) created in the collisions.
 The LHC at the CERN has studied proton-proton collisions at a
center-of-mass energy (per nucleon pair) of 7 TeV and heavy-ion
collisions at 2.76 TeV, and will study proton-proton collisions at
14 TeV~\cite{CMS:2008qya} and heavy-ion collisions at 5.5 TeV, which
are much higher than the maximum collision energy at the RHIC. As
the collision energy increases, a much broader and deeper study of
QGP will be done at the LHC. It leads to a significant extension of
the kinematic range in longitudinal rapidity and transverse
momentum. A systematic study of charged hadron multiplicities
$N_{ch}$
 is very important in understanding
the basic production mechanism of hadrons produced in
nucleon-nucleon and nucleus-nucleus collision experiments. Moreover,
the interpretation of heavy-ion results depends crucially on the
comparison with results from smaller collision systems such as
proton-proton ($pp$) and proton-nucleus
($p$A)~\cite{Abelev:2013haa}.

In recent years, some phenomenological models of initial-coherent
multiple interactions and particle transports~\cite{t1,t2} were
proposed and developed to explain the abundant experimental data.
But it is difficult to describe consistently the global properties
of final-state particles produced in high-energy reactions in a
single model. The bulk matter created in high-energy collisions can
be quantitatively described in terms of hydrodynamic and statistical
models~\cite{Andronic:2008gu, Cleymans:2006xj}, which are governed
mainly by the chemical freeze-out temperature and the baryochemical
potential. These models provide an accurate description of the data
over a large range of center-of-mass
energies~\cite{BraunMunzinger:2003zz}. With Tsallis statistics'
development and success in dealing with nonequilibrated complex
systems in condensed matter research, it has been used to understand
the particle production in high-energy physics. The Tsallis
statistics has been widely applied in the experpimental measurements
at RHIC~\cite{Adare:2011vy, Adare:2010fe, Abelev:2006cs} and
LHC~\cite{Aamodt:2010my, Aad:2010ac, Khachatryan:2010xs,
Khachatryan:2010us}. And it also has been discussed in literature,
e. g., Refs. \cite{Cleymans:2013rfq, Wong:2013sca}. In our previous
work~\cite{liu3}, the temperature information of emission sources
was understood indirectly  by a excitation degree, which varies with
location in a cylinder.  We have obtained emission source location
dependence of the exciting degree specifically. In this work, we
obtain the temperature of emission sources directly by parametrizing
experimentally measured $P_T$ spectra in Tsallis statistics. Then,
we reproduce the pseudorapidity distributions of
 identified particles produced in nucleon-nucleon and nucleus-nucleus
collision experiments.

This paper is organized as follows. In Sec. 2, we introduce the
Tsallis statistics in the multisource thermal model. In Sec. 3, we
present our numerical results, which are compared with the
experimental data in detail. At the end, we give conclusions in Sec.
4.

{\section{MODEL}}

Firstly, we embed the framework of  Tsallis statistics into the
geometrical picture of the multisource thermal model. In Tsallis
statistics, more than one version of the Tsallis distribution is
used to study the transverse distribution of identified particles
produced in high-energy collisions. Recently, an improved form of
the Tsallis distribution was proposed and it can naturally meet the
thermodynamic consistency~\cite{Cleymans:2012ya}. The quantum form
of the Tsallis distribution succeeded in describing the transverse
distribution measured by the ALICE and CMS collaborations. According
to the framework, the corresponding number of final-state particles
is given by \bea N=gV \int
\frac{d^3P}{(2\pi)^3}\left[1+(q-1)\frac{E-\mu}{T}\right
]^{-q/(q-1)},\label{b}
 \eea
where $P$, $E$, $T$, $\mu$, $V$, and $g$  are the momentum, the
energy, the temperature, the chemical potential,  the volume and the
degeneracy factor, respectively, and $q$ is a parameter
characterizing the degree of nonequilibrium. The momentum
distribution can be obtained as
 \bea E\frac{d^3N}{d^3P}=
\frac{gVE}{(2\pi)^3}\left[1+(q-1)\frac{E-\mu}{T}\right
]^{-q/(q-1)}.\label{b}
 \eea
When the parameter $q$ tends to 1, it is a standard Boltzmann
distribution. For zero chemical potential,  a transverse momentum
spectrum in terms of $y'$ (the particle rapidity in the rest frame
of a considered source) and $m_T$ (the transverse momentum) is given
by
 \bea \frac{d^2N}{dy'P_TdP_T}=
\frac{gVm_T\cosh{y'}}{(2\pi)^2}\left[1+(q-1)\frac{m_T\cosh{y'}}{T}\right
]^{-q/(q-1)}.\label{b}
 \eea
Then, we obtain a distribution function of the rapidity $y'$
 \bea f(y')=\frac{dN}{dy'}=gV\int
\frac{dP_T}{(2\pi)^2}m_TP_T\cosh{y'}\left[1+(q-1)\frac{m_T\cosh{y'}}{T}\right
]^{-q/(q-1)},\label{b}
 \eea
which is only the rapidity distribution of particles emitted in the
 emission source at a certain longitudinal location.
 In rapidity space ($y$ space), the longitudinal displacement of
the considered source needs to be taken into
account~\cite{Feng:2000pca}. Therefore, in a fixed emission source
with rapidity $y_x$ in the laboratory or center-of-mass reference
frame, the rapidity distribution of produced particles is given by
 \bea f(y, y_x)=gV\int
\frac{dP_T}{(2\pi)^2}m_TP_T\cosh{(y-y_x)}\left[1+(q-1)\frac{m_T\cosh{(y-y_x)}}{T}\right
]^{-q/(q-1)}.\label{b}
 \eea

Secondly, we briefly describe the geometrical picture of high-energy
collisions. In the  multisource thermal model~\cite{liu3,
Liu:2004rm} and nuclear geometry theory, a projectile cylinder and a
target cylinder are produced at $y$ space when the projectile and
target pass each other. In the laboratory reference system or center
of mass, we assume that the projectile cylinder is in the positive
rapidity direction and the target cylinder is in the negative one,
with rapidity ranges [$y_{pmin}$, $y_{pmax}$] and [$y_{tmin}$,
$y_{tmax}$], respectively. The projectile and target cylinder are
composed of  a series of isotropic emission sources with different
rapidity shifts. On both sides of the two cylinders, there are
leading particles appearing as two isotropic emission sources with
rapidity shifts $y_P$ and $y_T$, respectively. A thick
double-cylinder is formed in nucleus-nucleus collisions, and a thin
double-cylinder is formed in nucleon-nucleon collisions. With the
increasing direction of the rapidity coordinate scalar, we divide
the collision system into four parts, the target leading particles
(TL), target cylinder (TC), projectile cylinder (PC), and projectile
leading particles (PL), respectively. To give a clear picture for
understanding the definitions of the variables and parts, different
rapidity shifts for different parts in rapidity space are roughly
 shown in Fig. 1. The two cylinders may overlap completely or overlap
partly or may be separated. It is expected that the collision energy
corresponding to the situation of separation is higher than that of
overlap. The cylinder is not really a specific shape, but it  may be
understood to be a range of the rapidity of emission sources.

The normalized rapidity distribution can be written as
 \bea f(y) &
= & k_tf(y,
y_T)+\frac{K_t}{y_{tmax}-y_{tmin}}\int^{y_{tmax}}_{y_{tmin}}f(y,
y_t)d{y_t}\nonumber\\  && +
\frac{K_p}{y_{pmax}-y_{pmin}}\int^{y_{pmax}}_{y_{pmin}} f(y,
y_p)d{y_p}+k_pf(y, y_P). \label{c}
 \eea
where $k_t$, $K_t$, $K_p$ and $k_p$ are the contributions of TL, TC,
PC, and PL, respectively. $y_p$ and $y_t$ denote  the locations of
the emission sources in the TC and PC at $y$ space, respectively.
For symmetric collisions, $K_t=K_p=k$, $k_t=k_p=(1-2k)/2$,
$y_{pmax}=-y_{tmin}$, and $y_{pmin}=-y_{tmax}$, the normalized
rapidity distribution is rewritten as

\bea f(y)& =
&\frac{k}{y_{pmax}-y_{pmin}}\left[\int^{y_{pmax}}_{y_{pmin}}f(y,
y_p)d{y_p} +\int^{-y_{pmin}}_{-y_{pmax}}f(y,
y_t)d{y_t}\right]\nonumber\\  && + \frac{(1-2k)}{2}\left[f(y,
y_P)+f(y, -y_P)\right].\label{c}
 \eea

In the present work, the Monte Carlo method is used to calculate the
rapidity distribution. According to the different contribution
ratios, the emission sources distribute randomly in the PC range
[$y_{pmin}$, $y_{pmax}$],  TC range [$y_{tmin}$, $y_{tmax}$], PL
region, or TL region. In the final state, the rapidities of
particles produced in the two cylinders are given by
 \bea
y_{PC}=(y_{pmax}-y_{pmin})R_1+y_{pmin}+y',
\nonumber\\y_{TC}=(y_{pmax}-y_{pmin})R_2-y_{pmax}+y',\label{d}
 \eea
where  $R_1$ and $R_2$ are  random variables in interval [0,1]. The
 rapidities of leading particles are
 \bea
y_{PL}=y_P+y', \nonumber\\
y_{TL}=-y_P+y'.\label{d}
 \eea
In the above expressions, $y'$ is calculated by  using the Monte
Carlo calculation of Eq. (4). In the case of $P_T>>m_0$, the
rapidity $y$ and pseudorapidity  $\eta$ are approximately equal to
each other. But, the condition of $P_T>>m_0$ is not always
satisfied.  A conversion between  the pseudorapidity distribution
$\frac{dN}{d\eta}$ and the rapidity distribution $\frac{dN}{dy}$ is
 \bea \frac{dN}{d\eta}=\frac{P}{E}\frac{dN}{dy}=J(\eta,
\langle m\rangle/\langle p_T\rangle)\frac{dN}{dy}\,\,\,\,,\label{pr}
 \eea
where a Jacobian of the transformation is \bea J(\eta, \langle
m\rangle/\langle P_T\rangle)=\frac{\cosh{\eta}}{\sqrt{1+(\langle
m\rangle/\langle P_T\rangle)^2+\sinh^2{\eta}}}\,\,\,\,.\label{pr}
 \eea

{\section{COMPARISON WITH EXPERIMENTAL RESULTS}}

{\subsection{Proton-(anti)proton collisions}}

To identify the validity of the model and fix the temperature $T$
and the $q$, Figs. 2 and 3 show invariant yields of final-state
particles produced in inelastic (INEL) $pp$ collisions at
center-of-mass energy (per nucleon pair) $\sqrt{\mathrm{\it
s_{NN}}}=200$ GeV. The symbols are the experimental data of the
PHENIX Collaboration~\cite{Adare:2008ad, Adler:2006wg,
Adare:2010fe}. The solid lines are the results calculated by the
improved Tsallis distribution. The maximum value of the observed
$P_T$ reaches about 12.0 GeV/$c$. It can be seen that the results
agree well with the experimental data in the region. The $\chi^2$
per degree of freedom ($\chi^2/{\rm dof}$) testing provides
statistical indication of the most probable value of corresponding
parameters. The maximum value is 1.210, and the minimum value is
0.324. The parameter values are given in Table I. The parameters $T$
and $q$ are likely stable to be constant values because of the
scaling properties of the transverse momentum.

Figure 4 shows the pseudorapidity distributions of charged particles
produced in INEL $pp$ collisions at $\sqrt{\mathrm{\it s_{NN}}}=200$
and $410$ GeV with $|\eta|$ ranging from 0.10 to 5.30. The solid
circles with the error bars represent the experimental data measured
by the PHOBOS Collaboration~\cite{Alver:2010ck}. The solid lines are
our results, which are in good agreement with the experimental data.
Figure 5 shows the pseudorapidity distributions of charged particles
produced in INEL $pp$ (or $p\overline{p}$) collisions at
$\sqrt{\mathrm{\it s_{NN}}}=53$, 200, 546, and 900 GeV. The results
are compared with measurements made by the UA5 and ALICE
collaborations~\cite{Alner:1987wb, Aamodt:2010ft}. The corresponding
parameter values obtained by fitting the experimental data are given
in Table II with $\chi^2/{\rm dof}$. From the values, it is found
that the $y_{pmax}$ ($-y_{tmin}$), $y_{pmin}$ ($-y_{tmax}$) and
$y_P$ ($-y_T$) increase with increasing the collision energy. So,
the gap between the two cylinders and the length of each cylinder
also increase with increasing the collision energy. The parameter
$k$ does not change obviously, which means that the leading particle
contribution is almost identical. From the comparisons, we can see
that the multisource thermal model can describe the pseudorapidity
distributions of charged particles produced in INEL $pp$ (or
$p\overline{p}$) collisions over an energy range from 53 to 900 GeV
by using three rapidity shifts $y_{pmin}$, $y_{pmax}$ and $y_{P}$ as
free parameters.

The pseudorapidity distribution of charged particles produced in
INEL $p\overline{p}$ collisions at $\sqrt{\mathrm{\it s_{NN}}}=630$
GeV is presented in Fig. 6. The experimental data are measured by
the P238 Collaboration~\cite{Harr:1997sa}. The range of $|\eta|$ is
$1.5-5.5$. The parameter values taken in the calculation are given
in Table II with $\chi^2/{\rm dof}$. One can see that the
multisource thermal model with the three free parameters describes
the pseudorapidity distribution of charged particles produced in
INEL $p\overline{p}$ collisions at $\sqrt{\mathrm{\it s_{NN}}}=630$
GeV, which is higher than RHIC energies and less than LHC energies.
Figure 7 shows the pseudorapidity distributions of charged particles
produced in INEL $p\overline{p}$ collisions  at $\sqrt{\mathrm{\it
s_{NN}}}$=630 and 1800 GeV with $|\eta|$ ranging from 0 to 3.5. The
experimental data are measured by the CDF
Collaboration~\cite{Abe:1989td}. By fitting the data, the obtained
parameters are given in Table II with $\chi^2/{\rm dof}$. The
multisource thermal model
 can also describe the pseudorapidity
distribution of charged particles produced in INEL $p\overline{p}$
collisions at $\sqrt{\mathrm{\it s_{NN}}}=630$ and 1800 GeV.

For LHC energies, we present the pseudorapidity distributions of
charged particles produced in INEL $pp$ collisions at
$\sqrt{\mathrm{\it s_{NN}}}$=2.76 and 7 TeV with $|\eta|$ ranging
from 0.25 to 2.25 in Fig. 8. The experimental data are measured by
the CMS Collaboration~\cite{Khachatryan:2010us}. The obtained
parameters are given in Table II with $\chi^2/{\rm dof}$. Our
results are also in good agreement with the data.

From Table II, it is found that the three parameters obtained by the
comparison exhibit the linear dependences on
 ln$\sqrt{\mathrm{\it s_{NN}}}$. They are given in Fig. 9.
The symbols denote the parameters of different collaborations as
marked in the figure. The solid lines denote the fitted results, i.
e., $y_{pmin}=-y_{tmax}=(0.026\pm0.004)$ln$\sqrt{\mathrm{\it
s_{NN}}}$$+(0.020\pm0.011)$,
$y_{pmax}=-y_{tmin}=(0.506\pm0.041)$ln$\sqrt{\mathrm{\it
s_{NN}}}$$+(0.642\pm0.061)$ and
$y_{P}=-y_{T}=(0.211\pm0.047)$ln$\sqrt{\mathrm{\it
s_{NN}}}$$+(2.855\pm0.645)$. There is also a linear relationship
between the normalization coefficient $N_c$ and ln$\sqrt{\mathrm{\it
s_{NN}}}$, $N_c=(8.469\pm0.032)$ln$\sqrt{\mathrm{\it
s_{NN}}}$$-(26.788\pm1.570)$. The $\chi^2/{\rm dof}$ are 0.012,
0.040, 0.029 and 0.021, respectively. According to the linear laws,
the values of the parameters used in the model for
$pp$($p\overline{p}$) collisions at higher energies can be
predicted. Then, we may predict the pseudorapidity distributions of
charged particles produced at LHC energies. When $\sqrt{\mathrm{\it
s_{NN}}}$ rises up to 10 and 14 TeV,  the parameters are taken to be
$y_{pmin}=$ 0.260 and 0.271, $y_{pmax}$=5.297 and 5.471,
$y_{P}$=4.798 and 4.862, and $N_c$=51.280 and 54.204. The prediction
of the pseudorapidity distribution are given in Fig. 10.

{\subsection{Nucleus-nucleus collisions}}

 Figures 11 and 12 show the $P_T$ spectra of charged hadrons and
$\eta$ particles for the different centralities in Au+Au collisions
at $\sqrt{\mathrm{\it s_{NN}}}=200$ GeV with $|\eta|$ ranging from
0.1 to 5.3. The symbols are the experimental data from the PHOBOS
Collaboration~\cite{Back:2003qr} and the PHENIX
Collaboration~\cite{Adler:2006hu}. The solid lines are the results
fitted by the improved Tsallis distribution. The parameter values
are given in Table III with $\chi^2/{\rm dof}$. Because the scaling
properties of the transverse momentum, the values of $T$ and $q$ do
not change significantly with the centralities. The maximum value of
$\chi^2/{\rm dof}$ is 0.671, and the minimum value is 0.358.

The pseudorapidity distributions of charged particles for eleven
centrality bins in Au+Au collisions at $\sqrt{\mathrm{\it
s_{NN}}}=200$ GeV are presented in Fig. 13. The symbols with the
error bars denote the experimental data of the PHOBOS
Collaboration~\cite{Alver:2010ck}. The solid lines denote our
results. In the calculation, the parameters used in the calculation
are  given in Table IV with $\chi^2/{\rm dof}$. The $y_{pmax}$
 ($-y_{tmin}$), $y_{pmin}$ ($-y_{tmax}$) and $y_P$ ($-y_T$) increase
linearly with increasing the centrality or decrease linearly with
increasing the centrality percentage $C$. They are plotted in Fig.
14, where the solid lines are the fitted results
$y_{pmin}=-y_{tmax}=-0.001$$C$$+(0.110\pm0.021)$,
$y_{pmax}=-y_{tmin}=-(0.008\pm0.001)$$C$$+(3.574\pm0.042)$, and
$y_{P}=-y_{T}=-(0.016\pm0.002)$$C$$+(4.604\pm0.049)$. The
normalization coefficient is
 \bea
N_c=(6020.5\pm37.5)\exp{(-\frac{C}{0.305\pm0.006})}-(300.5\pm71.5).
\label{nc1}
 \eea
According to the laws, the pseudorapidity distributions of charged
particles for other centralities may be predicted.

Figure 15 shows the pseudorapidity distributions of charged
particles for twelve centrality bins in Cu+Cu collisions at
$\sqrt{\mathrm{\it s_{NN}}}=200$ GeV. The symbols and lines
represent the same meanings as those in Fig. 13. Our results are
also in good agreement with the experimental data. The parameters
are given in Table IV with $\chi^2/{\rm dof}$. By fitting the
parameters, the obtained relationships between the parameters and
$C$ are $y_{pmin}=-y_{tmax}=-0.001$$C$$+(0.115\pm0.006)$,
$y_{pmax}=-y_{tmin}=-(0.009\pm0.001)$$C$$+(3.514\pm0.039)$,
$y_{P}=-y_{T}=-(0.014\pm0.002)$$C$$+(4.460\pm0.055)$, and the
normalization coefficient is
$N_c=(1820.7\pm11.5)\exp{(-\frac{C}{0.355\pm0.005})}-(190.5\pm39.5)$
  as given in Fig. 16.

Figure 17 shows the pseudorapidity distributions of charged
particles for seven centrality bins in Pb+Pb collisions at
$\sqrt{\mathrm{\it s_{NN}}}=2.76$ TeV. The symbols with the error
bars are the experimental data measured at the
LHC~\cite{Abbas:2013bpa}. The solid lines are our results.  From the
figure, we know that the multisource thermal model with the three
free parameters can also describe the pseudorapidity distributions
of charged particles produced in Pb+Pb collisions at
$\sqrt{\mathrm{\it s_{NN}}}=2.76$ TeV. The parameter values are
given in Table IV with $\chi^2/{\rm dof}$. The parameters as the
function of $C$ are
$y_{pmin}=-y_{tmax}=-0.001$$C$$+(0.129\pm0.007)$,
$y_{pmax}=-y_{tmin}=-(0.009\pm0.001)$$C$$+(3.982\pm0.045)$,
$y_{P}=-y_{T}=-(0.010\pm0.002)$$C$$+(5.080\pm0.057)$, and the
normalization coefficient is
$N_c=(17580.5\pm104.4)\exp{(-\frac{C}{0.251\pm0.004})}-(600.5\pm112.5)$
  as given in Fig. 18.

{\section{CONCLUSIONS}}

We embed consistently the improved form of the Tsallis distribution
into the multisource thermal model for describing hadron productions
in $pp$ (or $p\overline{p}$) and $AA$ collisions at RHIC and LHC
energies. The pseudorapidity distributions have been systematically
investigated and compared to the experimental data for
$pp$($\overline{p}$) and Au+Au, Cu+Cu, and Pb+Pb collisions at the
RHIC and LHC energies. The updated multisource thermal model can
describe the experimental results. The three free parameters
$y_{pmax}$ ($-y_{tmin}$), $y_{pmin}=-y_{tmax}$, and $y_P$ (-$y_T$)
  taken in the calculations exhibit certain regularities for the
collision energy and the  collision centrality. The linear
dependences of the parameters on ln$\sqrt{\mathrm{\it s_{NN}}}$ are
found. It may be used to predict the pseudorapidity distributions of
produced particles in $pp$($\overline{p}$) at higher colliding
energies such as LHC energies. And it is also used to predict the
pseudorapidity distributions of produced particles in Au+Au, Cu+Cu,
and Pb+Pb collisions with other centralities at high energies. As an
example, we have given the predictions of the pseudorapidity
distributions of charged particles produced in $pp$($\overline{p}$)
collisions at higher energies.

The multisource thermal model was developed by us  in the past
years~\cite{liu3, Liu:2004rm}. This model assumes that many emission
sources of produced particles and nuclear fragments are formed in
high-energy collisions. The particles are emitted isotropically in
the rest frame of the emission sources with the different excitation
degree in collisions. Each emission source was treated as a thermal
equilibrium system of classical ideal gas.  So, the classical
Maxwell¡¯s ideal gas distribution was adopted without considering
the effects of the relativity and quantum. Recently, the improved
Tsallis distribution~\cite{Cleymans:2012ya} was suggested in the
particular case of relativistic high-energy quantum distributions.
Moreover, the thermodynamic consistency of the distribution was
considered in detail. The temperature $T$ and the degree of
nonequilibrium $q$ can be determined by a global fit of the $P_T$
spectra. In the present work, we combined the improved Tsallis
distribution and the picture of the multisource thermal model. The
pseudorapidity distributions are mainly related to the rapidity of
the emission sources in the formalism.

In summary, the pseudorapidity distributions of charged particles
produced in nucleus-nucleus and nucleon-nucleon  collisions at
 RHIC and LHC energies have been studied in the improved multisource thermal
 model, where the improved Tsallis distribution is embedded.
 The results in each collision are compared with
experimental data measured by different collaborations. Our
investigations indicate the improved model is successful in the
description of hadron productions. At the same time, it is found
 that the rapidity shifts of the two cylinders
are linearly related to ln$\sqrt{\mathrm{\it s_{NN}}}$. According to
the laws, the predictions of the results at higher-energy collisions are given. \\

{\bf Acknowledgments.} This work is supported by the National
Natural Science Foundation of China under Grants No. 11247250, No.
11005071, and No. 10975095; the National Fundamental Fund of
Personnel Training under Grant No. J1103210, the Shanxi Provincial
Natural Science Foundation under Grants No. 2013021006 and No.
2011011001; the Open Research Subject of the Chinese Academy of
Sciences Large-Scale Scientific Facility under Grant No. 2060205,
and the Shanxi Scholarship Council of China.

\newpage

\begin{table}[ht]
\vspace*{-0.0cm} {\small  \vspace*{-0.1cm} \caption{Parameter values corresponding to the solid curves in Fig. 2 and Fig. 3.}%
\label{table-1}%
\vspace*{0.0cm}
\begin{center}{\begin{tabular}{c c c c c c} \hline\hline
\small Figure & Collision & Particle  & $T$ (GeV)  & $q$   & $\chi^2$/dof \\
\hline
2(a) & $pp$               & $\pi^0$              & $0.047\pm0.005$   & $1.077\pm0.015$  & $0.324$\\
2(b) & $pp$               & $\eta$               & $0.047\pm0.005$   & $1.082\pm0.010$  & $0.390$\\
3(a) & $p\overline{p}$    & $\omega$             & $0.047\pm0.005$   & $1.082\pm0.012$  & $1.210$\\
     & $p\overline{p}$    & $K_s^0$              & $0.047\pm0.005$   & $1.078\pm0.015$  & $0.675$\\
     & $pp$               & $\eta'$              & $0.047\pm0.005$   & $1.082\pm0.010$  & $0.624$\\
     & $pp$               & $\phi$               & $0.047\pm0.005$   & $1.082\pm0.009$  & $0.647$\\
3(b) & $p\overline{p}$    & $\pi$                & $0.050\pm0.005$   & $1.079\pm0.012$  & $0.546$\\
     & $p\overline{p}$    & $K$                  & $0.050\pm0.005$   & $1.079\pm0.010$  & $0.575$\\
     & $p\overline{p}$    & $\eta$               & $0.050\pm0.005$   & $1.079\pm0.010$  & $0.951$\\
     & $p\overline{p}$    &  $\omega$            & $0.050\pm0.005$   & $1.082\pm0.015$  & $0.744$\\
     & $p\overline{p}$    & $p(\overline{p})$    & $0.050\pm0.005$   & $1.082\pm0.015$  & $0.524$\\
     & $pp$               & $\eta'$              & $0.050\pm0.005$   & $1.082\pm0.012$  & $0.803$\\
     & $pp$               &  $\phi$              & $0.050\pm0.005$   & $1.082\pm0.012$  & $1.112$\\
\hline\hline
\end{tabular}}

\end{center}}
\end{table}

\begin{table}[ht]
\vspace*{-0.0cm} {\small  \vspace*{-0.1cm} \caption{Parameter values corresponding to the solid curves in Figs. 4-8.}%
\label{table-2}%
\vspace*{0.0cm}
\begin{center}{\begin{tabular}{c c c c c c c c c } \hline\hline
\small Figure & Energy (GeV)  & Collision&  $y_{pmax}$ & $y_{pmin}$ & $y_P$ & $k$ & $N_c$ & $\chi^2$/dof \\
\hline
4(a) & 200      & $pp$             & $3.35\pm0.08$     & $0.162\pm0.008$   & $3.94\pm0.08$       & $0.432\pm0.010$    & $18.20\pm0.85$  & $0.915$\\
4(b) & 410      & $pp$             & $3.65\pm0.09$     & $0.179\pm0.007$   & $4.08\pm0.07$       & $0.432\pm0.010$    & $24.65\pm0.94$  & $0.726$\\
5(a) & 53       & $p\overline{p}$  & $2.67\pm0.05$     & $0.123\pm0.005$   & $3.71\pm0.05$       & $0.432\pm0.010$    &  $7.24\pm0.70$  & $1.127$\\
5(b) & 200      & $p\overline{p}$  & $3.31\pm0.07$     & $0.165\pm0.007$   & $3.88\pm0.07$       & $0.432\pm0.010$    & $17.60\pm0.75$  & $0.958$\\
5(c) & 546      & $p\overline{p}$  & $3.85\pm0.10$     & $0.188\pm0.006$   & $4.17\pm0.08$       & $0.432\pm0.010$    & $25.80\pm0.91$  & $0.549$\\
5(d) & 900      & $pp$ and $p\overline{p}$  & $4.05\pm0.11$     & $0.200\pm0.008$   & $4.30\pm0.09$       & $0.432\pm0.010$    & $30.50\pm1.05$  & $0.752$\\
6    & 630      & $p\overline{p}$  & $3.94\pm0.10$     & $0.192\pm0.005$   & $4.29\pm0.08$       & $0.432\pm0.010$    & $27.80\pm0.95$  & $1.265$\\
7(a) & 630      & $p\overline{p}$  & $3.99\pm0.12$     & $0.194\pm0.006$   & $4.16\pm0.06$       & $0.432\pm0.010$    & $28.50\pm1.00$  & $0.650$\\
7(b) & 900      & $p\overline{p}$  & $4.41\pm0.13$     & $0.221\pm0.008$   & $4.36\pm0.10$       & $0.432\pm0.010$    & $37.20\pm1.15$  & $0.461$\\
8(a) & 2360     & $pp$             & $4.56\pm0.14$     & $0.225\pm0.009$   & $4.48\pm0.12$       & $0.432\pm0.010$    & $39.56\pm1.24$  & $0.650$\\
8(b) & 7000     & $pp$             & $5.15\pm0.15$     & $0.254\pm0.011$   & $4.76\pm0.14$       & $0.432\pm0.010$    & $48.46\pm1.47$  & $0.461$\\
\hline\hline
\end{tabular}}

\end{center}}
\end{table}

\begin{table}[h]
\vspace*{-0.0cm} {\small  \vspace*{-0.1cm} \caption{Parameter values corresponding to the solid curves in Fig. 11 and Fig. 12.}%
\label{table-3}%
\vspace*{0.0cm}
\begin{center}{\begin{tabular}{c c c c c} \hline\hline
\small Figure &  Centrality  & $T$ (GeV)  & $q$   & $\chi^2$/dof \\
\hline
11    & $0-6\%$             & $0.065\pm0.005$    & $1.069\pm0.004$   & $0.502$\\
     & $6-15\%$             & $0.065\pm0.005$    & $1.069\pm0.004$   & $0.655$\\
     & $15-25\%$            & $0.065\pm0.005$    & $1.069\pm0.004$   & $0.618$\\
     & $25-35\%$            & $0.065\pm0.005$    & $1.069\pm0.004$   & $0.560$\\
     & $35-45\%$            & $0.065\pm0.005$    & $1.069\pm0.004$   & $0.582$\\
     & $45-50\%$            & $0.065\pm0.005$    & $1.069\pm0.004$   & $0.671$\\
12    & $0-20\%$            & $0.060\pm0.005$    & $1.074\pm0.005$   & $0.358$\\
     & $20-60\%$            & $0.060\pm0.005$    & $1.074\pm0.005$   & $0.470$\\
     & $60-92\%$            & $0.060\pm0.005$    & $1.074\pm0.005$   & $0.502$\\
     & $0-92\%$             & $0.060\pm0.005$    & $1.074\pm0.005$   & $0.516$\\
\hline\hline
\end{tabular}}
\end{center}}
\end{table}

\begin{table}[th]
\vspace*{-0.1cm} {\small  \vspace*{-0.0cm} \caption{Parameter values corresponding to the solid curves in Fig. 13, Fig. 15 and Fig. 17.}%
\label{table-4}%
\vspace*{0.0cm}
\begin{center}{\begin{tabular}{c c c c c c c c c } \hline\hline
\small Figure & Collision  & Centrality &  $y_{pmax}$ & $y_{pmin}$ & $y_P$ & $k$ & $Nc$ & $\chi^2$/dof \\
\hline
13       & Au+Au   & $0-3\%$     & $3.57\pm0.07$  & $0.110\pm0.005$   & $4.61\pm0.07$ & $0.430\pm0.005$   & $5290\pm264$   & $0.550$\\
         &         & $3-6\%$     & $3.54\pm0.06$  & $0.106\pm0.005$   & $4.54\pm0.06$ & $0.430\pm0.005$   & $4895\pm245$   & $0.652$\\
         &         & $6-10\%$    & $3.49\pm0.05$  & $0.102\pm0.004$   & $4.48\pm0.05$ & $0.430\pm0.005$   & $4341\pm217$   & $0.484$\\
         &         & $10-15\%$   & $3.47\pm0.05$  & $0.097\pm0.004$   & $4.42\pm0.05$ & $0.430\pm0.005$   & $3763\pm188$   & $0.426$\\
         &         & $15-20\%$   & $3.44\pm0.04$  & $0.091\pm0.004$   & $4.36\pm0.04$ & $0.430\pm0.005$   & $3153\pm158$   & $0.424$\\
         &         & $20-25\%$   & $3.39\pm0.04$  & $0.086\pm0.004$   & $4.27\pm0.04$ & $0.430\pm0.005$   & $2645\pm132$   & $0.427$\\
         &         & $25-30\%$   & $3.36\pm0.05$  & $0.079\pm0.005$   & $4.18\pm0.04$ & $0.430\pm0.005$   & $2184\pm109$   & $0.520$\\
         &         & $30-35\%$   & $3.31\pm0.03$  & $0.073\pm0.004$   & $4.10\pm0.04$ & $0.427\pm0.005$   & $1819\pm91$    & $0.576$\\
         &         & $35-40\%$   & $3.27\pm0.02$  & $0.068\pm0.002$   & $4.01\pm0.05$ & $0.427\pm0.005$   & $1486\pm74$    & $0.583$\\
         &         & $40-45\%$   & $3.25\pm0.04$  & $0.064\pm0.003$   & $3.94\pm0.04$ & $0.427\pm0.005$   & $1204\pm60$    & $0.658$\\
         &         & $45-50\%$   & $3.21\pm0.04$  & $0.060\pm0.002$   & $3.90\pm0.02$ & $0.427\pm0.005$   & $951\pm48$     & $0.642$\\
15       & Cu+Cu   & $0-3\%$     & $3.51\pm0.06$  & $0.114\pm0.004$   & $4.45\pm0.05$ & $0.430\pm0.005$   & $1541\pm70$    & $0.572$\\
         &         & $3-6\%$     & $3.48\pm0.05$  & $0.110\pm0.005$   & $4.40\pm0.05$ & $0.430\pm0.005$   & $1407\pm68$    & $0.645$\\
         &         & $6-10\%$    & $3.44\pm0.05$  & $0.107\pm0.004$   & $4.35\pm0.05$ & $0.430\pm0.005$   & $1262\pm59$    & $0.506$\\
         &         & $10-15\%$   & $3.40\pm0.05$  & $0.102\pm0.002$   & $4.29\pm0.05$ & $0.430\pm0.005$   & $1084\pm51$    & $0.521$\\
         &         & $15-20\%$   & $3.37\pm0.04$  & $0.097\pm0.002$   & $4.22\pm0.04$ & $0.427\pm0.005$   & $917\pm43$     & $0.558$\\
         &         & $20-25\%$   & $3.31\pm0.04$  & $0.091\pm0.004$   & $4.18\pm0.04$ & $0.427\pm0.005$   & $771\pm38$     & $0.515$\\
         &         & $25-30\%$   & $3.26\pm0.04$  & $0.088\pm0.004$   & $4.09\pm0.06$ & $0.427\pm0.005$   & $645\pm32$     & $0.570$\\
         &         & $30-35\%$   & $3.22\pm0.03$  & $0.083\pm0.005$   & $4.01\pm0.04$ & $0.427\pm0.005$   & $538\pm27$     & $0.566$\\
         &         & $35-40\%$   & $3.19\pm0.03$  & $0.078\pm0.002$   & $3.95\pm0.05$ & $0.427\pm0.005$   & $445\pm23$     & $0.562$\\
         &         & $40-45\%$   & $3.15\pm0.04$  & $0.074\pm0.003$   & $3.88\pm0.04$ & $0.425\pm0.005$   & $364\pm19$     & $0.597$\\
         &         & $45-50\%$   & $3.11\pm0.03$  & $0.072\pm0.002$   & $3.83\pm0.05$ & $0.425\pm0.005$   & $293\pm15$     & $0.620$\\
         &         & $50-55\%$   & $3.08\pm0.03$  & $0.070\pm0.003$   & $3.80\pm0.05$ & $0.425\pm0.005$   & $234\pm13$     & $0.625$\\
17       & Pb+Pb   & $0-5\%$     & $3.96\pm0.05$  & $0.126\pm0.004$   & $5.06\pm0.04$ & $0.422\pm0.005$   & $15200\pm400$  & $0.925$\\
         &         & $5-10\%$    & $3.91\pm0.04$  & $0.123\pm0.002$   & $5.02\pm0.03$ & $0.422\pm0.005$   & $12350\pm240$  & $0.805$\\
         &         & $10-20\%$   & $3.84\pm0.03$  & $0.114\pm0.004$   & $4.97\pm0.03$ & $0.422\pm0.005$   & $9200\pm180$   & $0.778$ \\
         &         & $20-30\%$   & $3.75\pm0.05$  & $0.108\pm0.004$   & $4.88\pm0.04$ & $0.419\pm0.005$   & $6080\pm140$   & $1.156$\\
         &         & $30-40\%$   & $3.68\pm0.04$  & $0.096\pm0.002$   & $4.80\pm0.05$ & $0.419\pm0.005$   & $3950\pm80$    & $0.718$\\
         &         & $40-50\%$   & $3.59\pm0.03$  & $0.089\pm0.003$   & $4.72\pm0.04$ & $0.419\pm0.005$   & $2405\pm68$    & $0.627$\\
         &         & $50-60\%$   & $3.53\pm0.02$  & $0.083\pm0.002$   & $4.67\pm0.02$ & $0.419\pm0.005$   & $1340\pm46$    & $0.644$\\
\hline\hline
\end{tabular}}

\end{center}}
\end{table}

\clearpage

\begin{figure}[htbp]
\vskip13.5cm
\begin{center}
\includegraphics[width=0.75\textwidth] {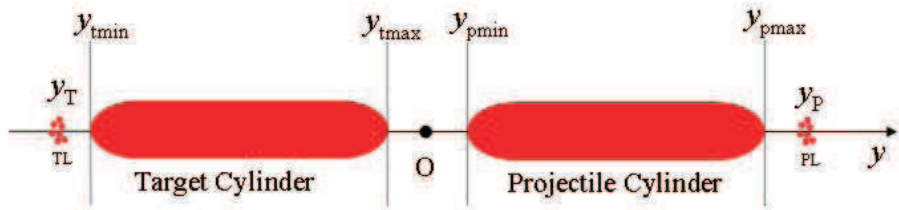}
\end{center}
\vskip-0.5cm \caption{(Color online) Schematic sketch of the
distribution of emission sources in rapidity space.}
\end{figure}

\begin{figure}[htbp]
\begin{center}
\vskip -0.cm
\includegraphics[width=0.75\textwidth]{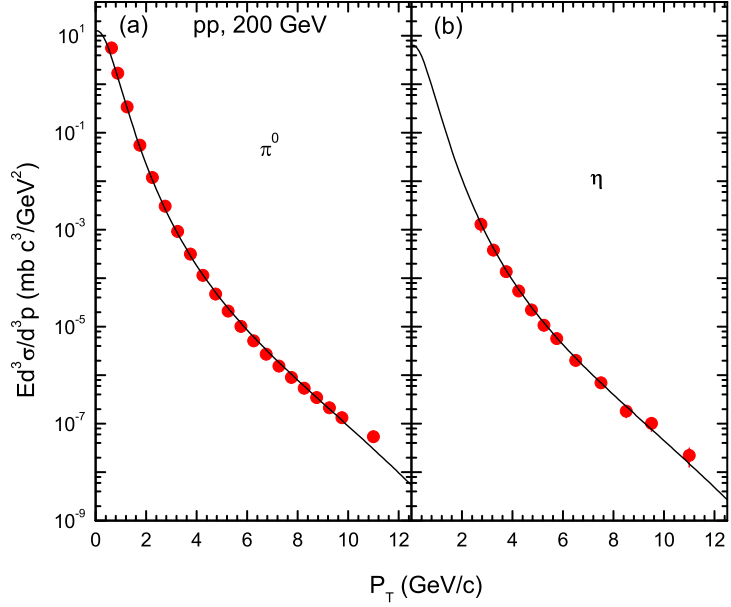}
\end{center} \vskip-0.7cm
 \caption{(Color online) Invariant $\pi^0$ and $\eta$ cross section as a
function of transverse momentum in $pp$ collisions at
$\sqrt{\mathrm{\it s_{NN}}}$ = 200 GeV.  Experimental data from the
PHENIX Collaboration~\cite{Adare:2008ad, Adler:2006wg} are shown by
the scattered symbols. Our calculated results are shown by the
curves.}
\end{figure}

\begin{figure}[htbp]
\begin{center}
\vskip 0.7cm
\includegraphics[width=0.75\textwidth]{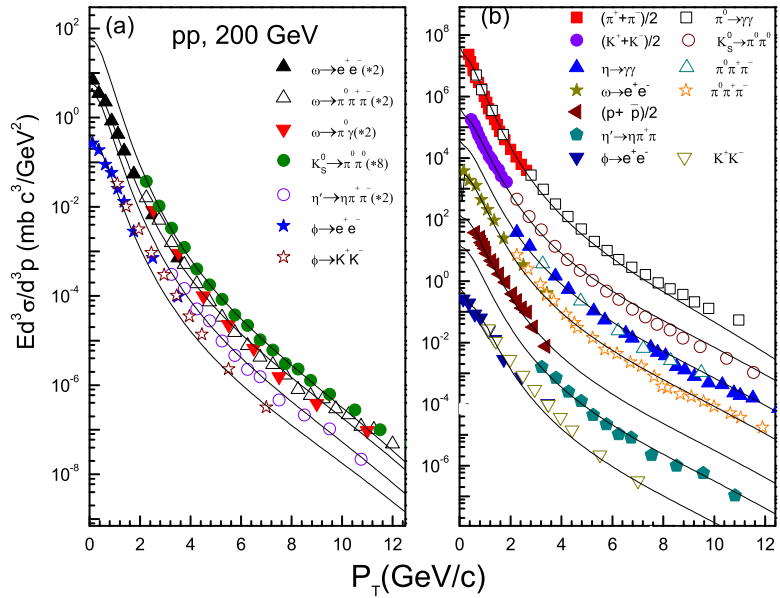}
\end{center} \vskip-1.1cm
 \caption{(Color online) Invariant differential cross section of  different
particles measured in $pp$ collisions at $\sqrt{\mathrm{\it
s_{NN}}}$ = 200 GeV in various decay modes. Experimental data
measured by the PHENIX Collaboration~\cite{Adare:2010fe} are shown
by the scattered symbols. The curves are our calculated results.}
\end{figure}

\begin{figure}[htbp]
\begin{center}
\vskip -0.cm
\includegraphics[width=0.80\textwidth]{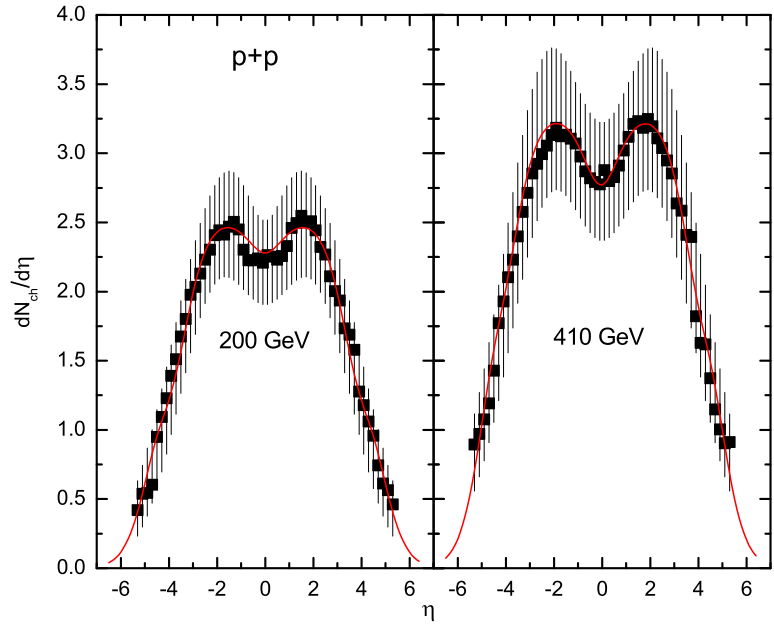}
\end{center} \vskip-1.1cm
 \caption{(Color online) The charged particle multiplicity $dN_{ch}/d\eta$ in $pp$ inelastic
collisions at $\sqrt{\mathrm{\it s_{NN}}}$ = 200 GeV and 410 GeV.
The symbols represent the experimental data of the PHOBOS
Collaboration~\cite{Alver:2010ck}.
 The curves are our calculated results.}
\end{figure}

\begin{figure}[htbp]
\begin{center}
\vskip -0.cm
\includegraphics[width=0.80\textwidth]{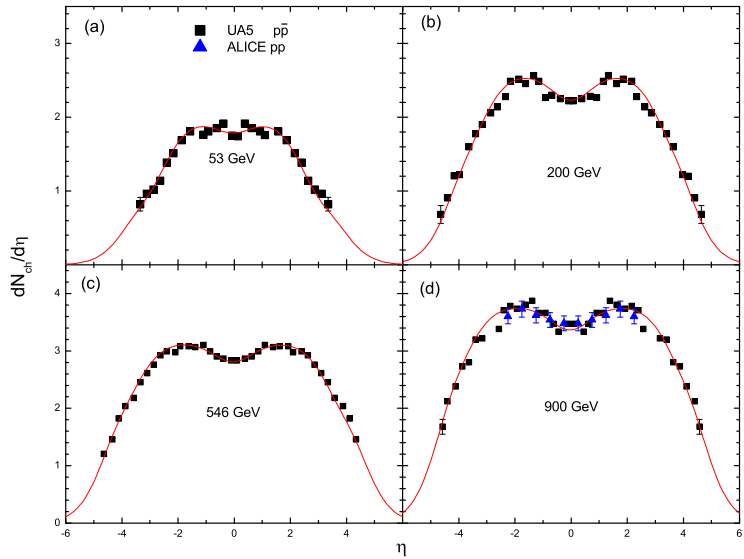}
\end{center} \vskip-1.7cm
 \caption{(Color online) The charged particle multiplicity $dN_{ch}/d\eta$
 at $\sqrt{\mathrm{\it s_{NN}}}$ = 53, 200, 546 and 900 GeV. The symbols
 represent the measured data from the UA5
Collaboration~\cite{Alner:1987wb} ($p\overline{p}$ collisions, with
statistical errors only) and the ALICE Collaboration~\cite{
Aamodt:2010ft} ($pp$  collisions, with statistical errors only). The
curves are our calculated results.}
\end{figure}

\begin{figure}[htbp]
\begin{center}
\includegraphics[width=0.85\textwidth]{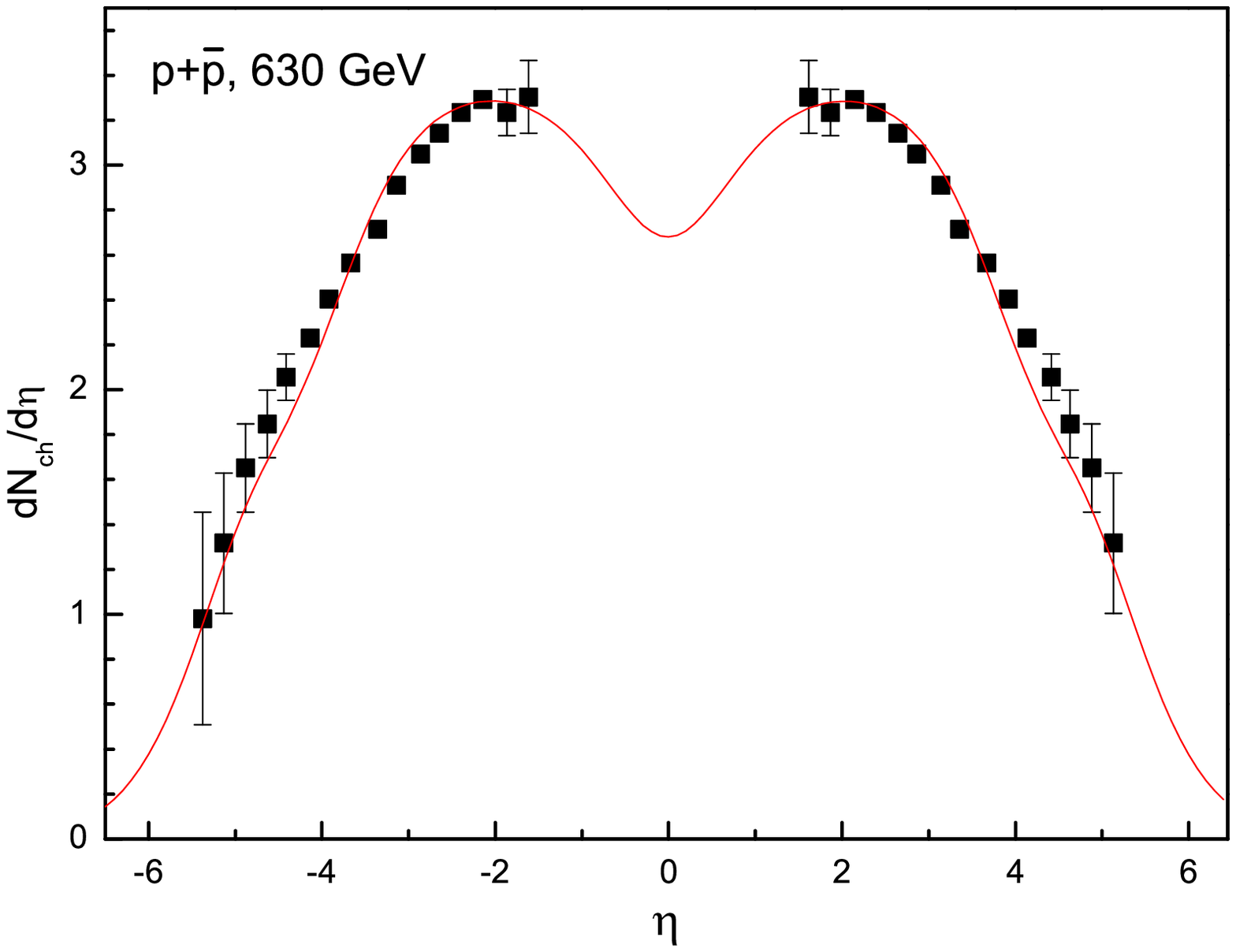}
\end{center} \vskip-1.cm
\caption{(Color online) The charged particle multiplicity
$dN_{ch}/d\eta$ in $p\overline{p}$ inelastic collisions at
$\sqrt{\mathrm{\it s_{NN}}}$ = 630 GeV. The symbols and curve
represent the same meanings as those in Fig. 5, but the experimental
data  are taken from the P238 Collaboration~\cite{Harr:1997sa}.}
\end{figure}

\begin{figure}[htbp]
\begin{center}
\includegraphics[width=0.85\textwidth]{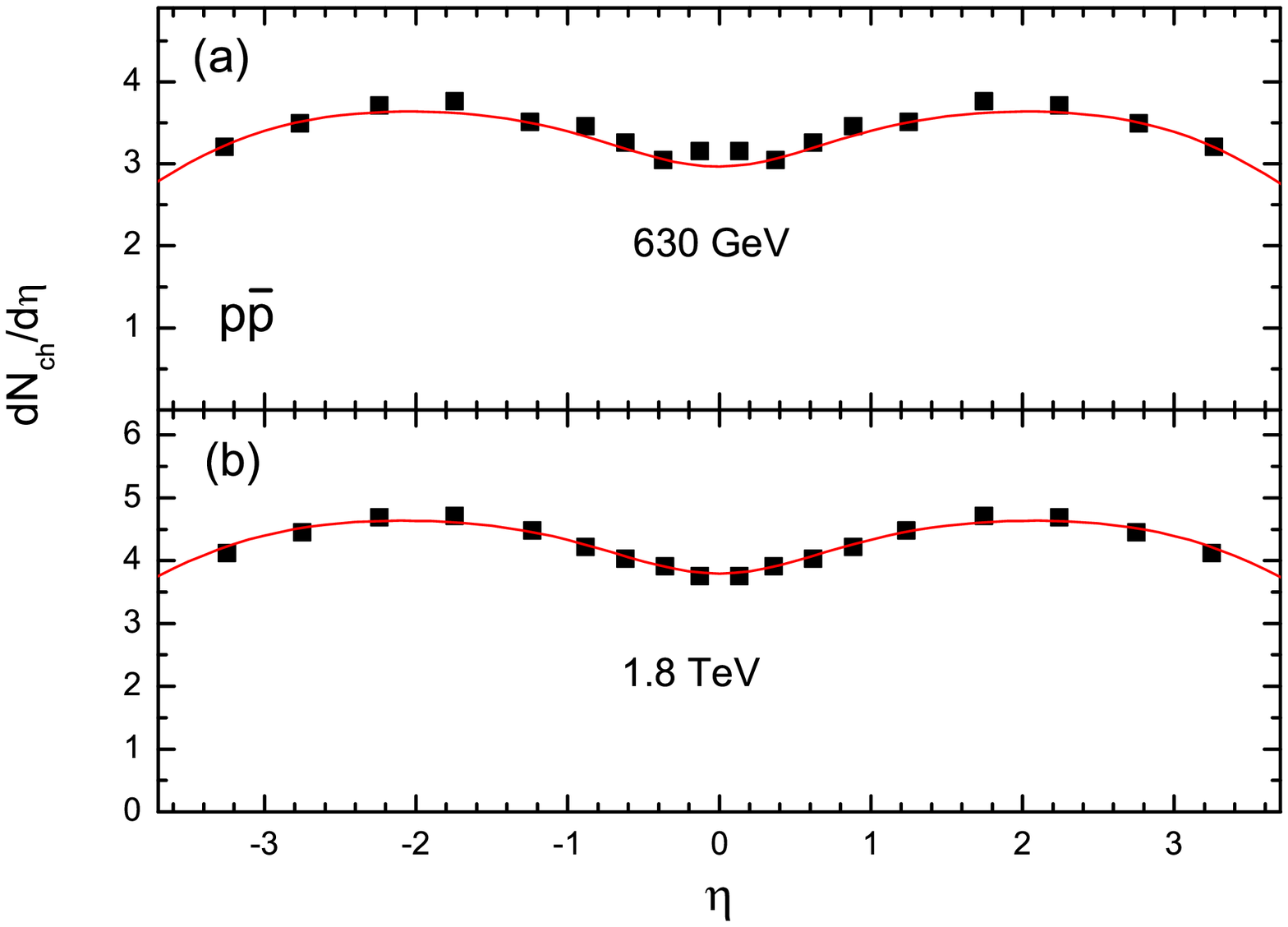}
\end{center} \vskip-1.5cm
\caption{(Color online) The charged particle multiplicity
$dN_{ch}/d\eta$ in $p\overline{p}$ inelastic collisions at
$\sqrt{\mathrm{\it s_{NN}}}$ = 630 GeV and 1.8 TeV. The symbols and
curves represent the same meanings as those in Fig. 5, but the
experimental data are taken from the CDF
Collaboration~\cite{Abe:1989td}.}
\end{figure}

\begin{figure}[htbp]
\begin{center}
\includegraphics[width=0.85\textwidth]{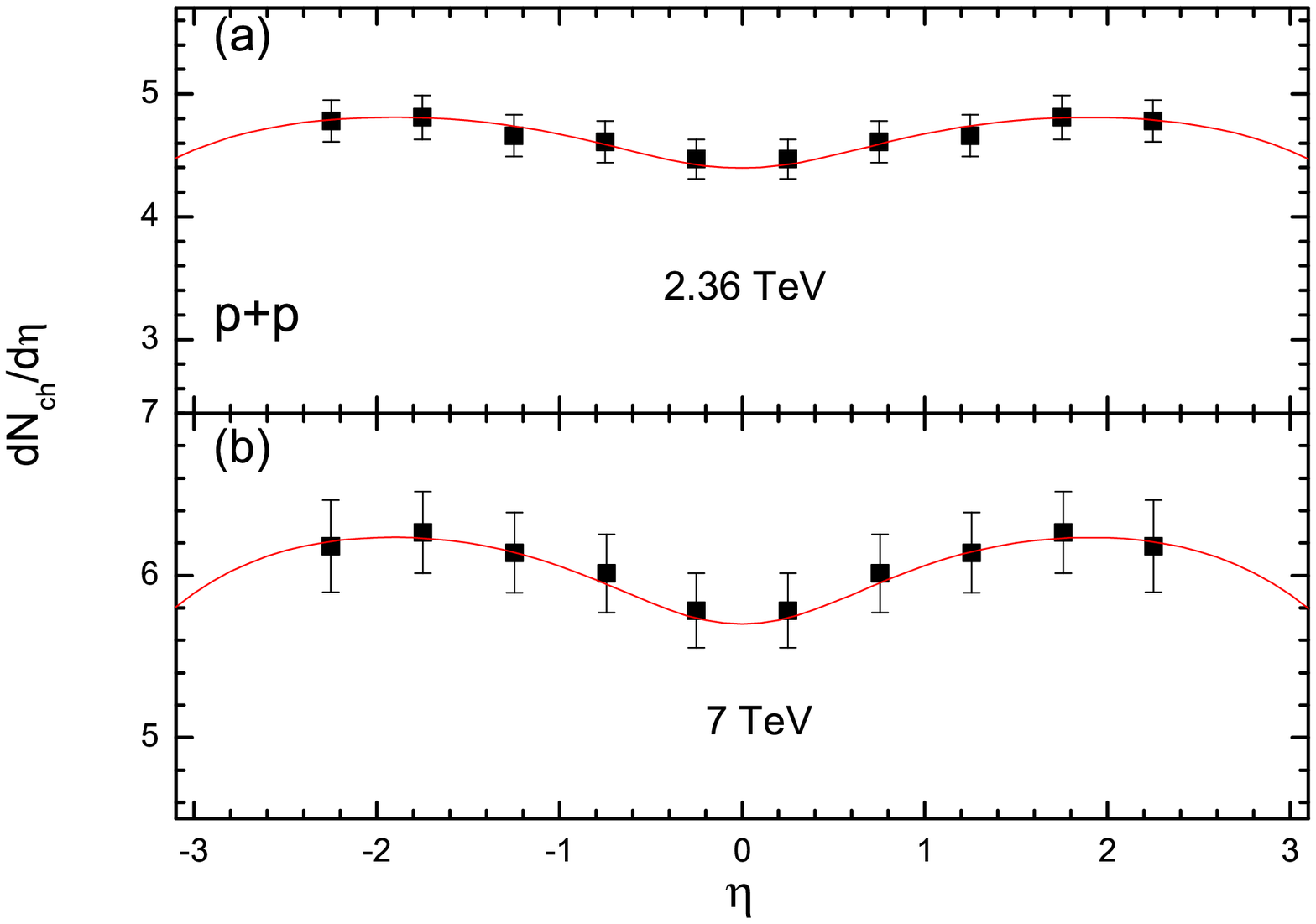}
\end{center} \vskip-1.5cm
\caption{(Color online) The charged particle multiplicity
$dN_{ch}/d\eta$ in $pp$ inelastic collisions at $\sqrt{\mathrm{\it
s_{NN}}}=$2.36 TeV and 7 TeV. The symbols and curves represent the
same meanings as those in Fig. 5, but the experimental data are
taken from the CMS Collaboration~\cite{Khachatryan:2010us}.}
\end{figure}

\begin{figure}[htbp]
\begin{center}
\includegraphics[width=0.85\textwidth]{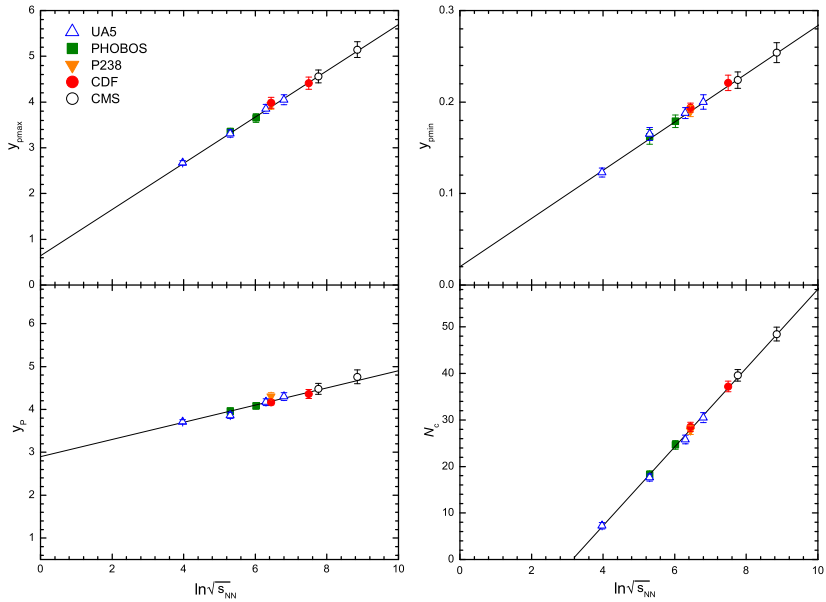}
\end{center} \vskip-1.5cm
\caption{(Color online) The dependence of the different parameters
on $\ln{\sqrt{\mathrm{\it s_{NN}}}}$. The symbols represent the
 parameter values used in the calculations for different experimental
collaborations.  The solid lines denote the fitted results. }
\end{figure}

\begin{figure}[htbp]
\begin{center}
\includegraphics[width=0.85\textwidth]{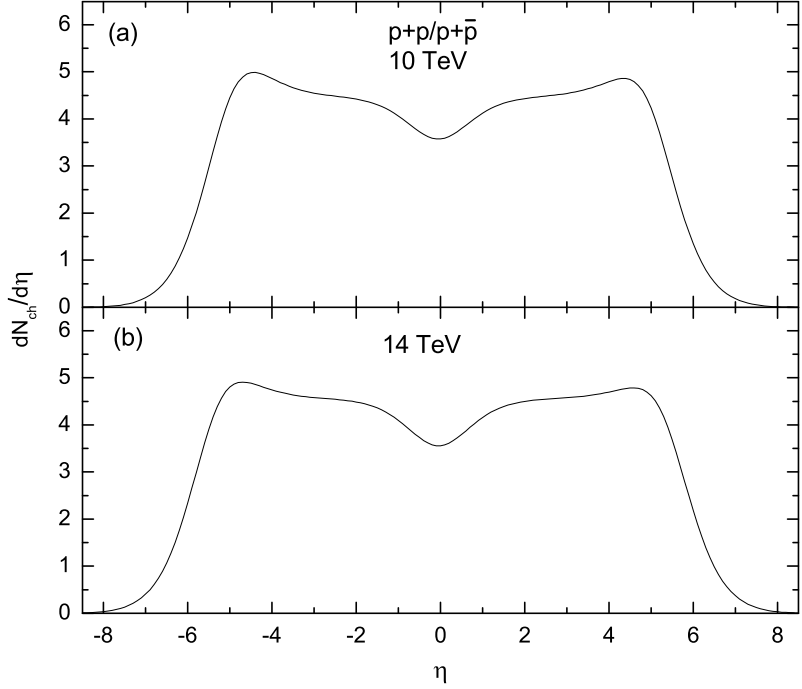}
\end{center} \vskip-1.0cm
\caption{(Color online) The charged particle multiplicity
$dN_{ch}/d\eta$ in $pp$ (or $p\overline{p}$)  collisions at
$\sqrt{\mathrm{\it s_{NN}}}$ = 10 TeV and 14 TeV.}
\end{figure}

\begin{figure}[htbp]
\begin{center}
\includegraphics[width=0.85\textwidth]{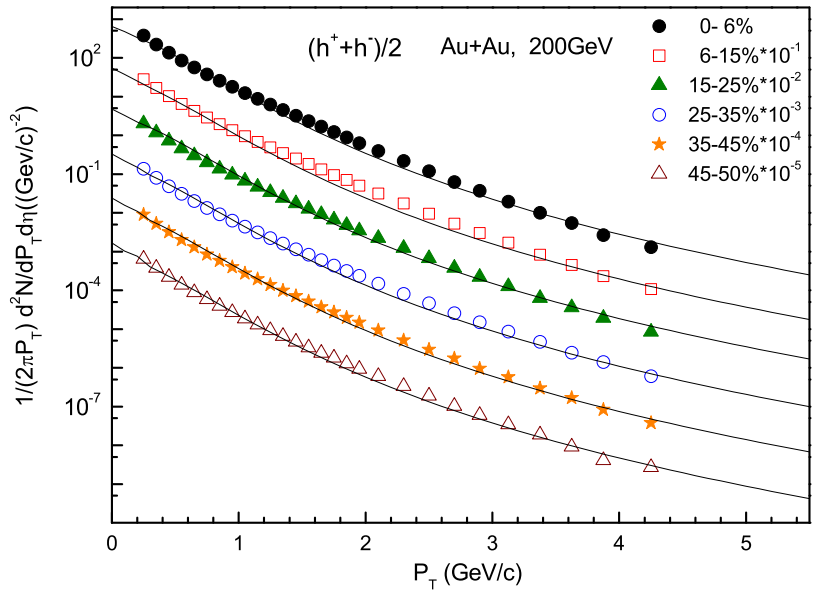}
\end{center} \vskip-1.cm
\caption{(Color online) Charged hadron transverse momentum
distributions in Au+Au collisions at $\sqrt{\mathrm{\it s_{NN}}}$
=200 GeV. For clarity, consecutive bins are
scaled by factors of 10.  
Experimental data of the PHOBOS Collaboration~\cite{Back:2003qr} are
shown by the scattered symbols. Our calculated results are shown by
the curves.}
\end{figure}

\begin{figure}[htbp]
\begin{center}
\includegraphics[width=0.85\textwidth]{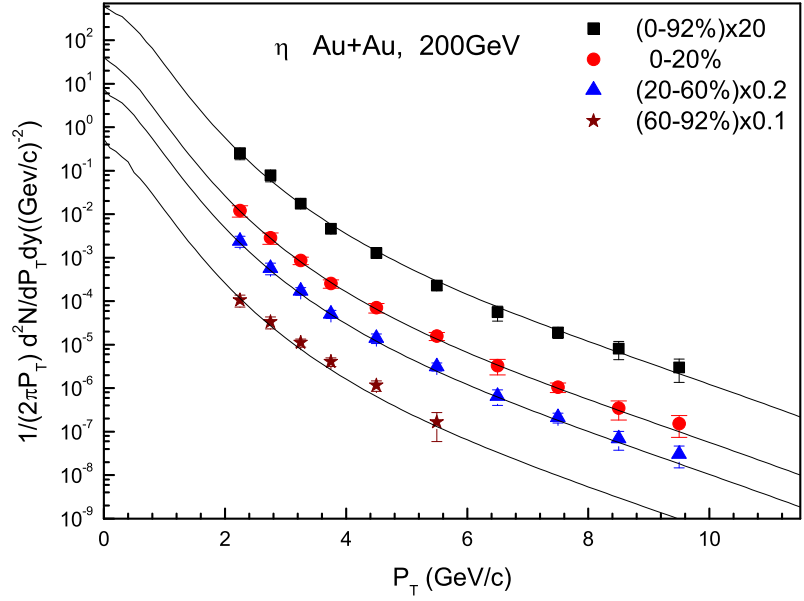}
\end{center} \vskip-1.cm
\caption{(Color online) $\eta$ transverse momentum distributions  in
Au+Au collisions at $\sqrt{\mathrm{\it s_{NN}}}$ = 200 GeV. The
error bars are the quadratic sum of statistical and systematic
uncertainties. Experimental data of the PHENIX
Collaboration~\cite{Adler:2006hu} are shown by the scattered
symbols. Our calculated results are shown by the curves.}
\end{figure}

\begin{figure}[htbp]
\begin{center}
\includegraphics[width=0.85\textwidth]{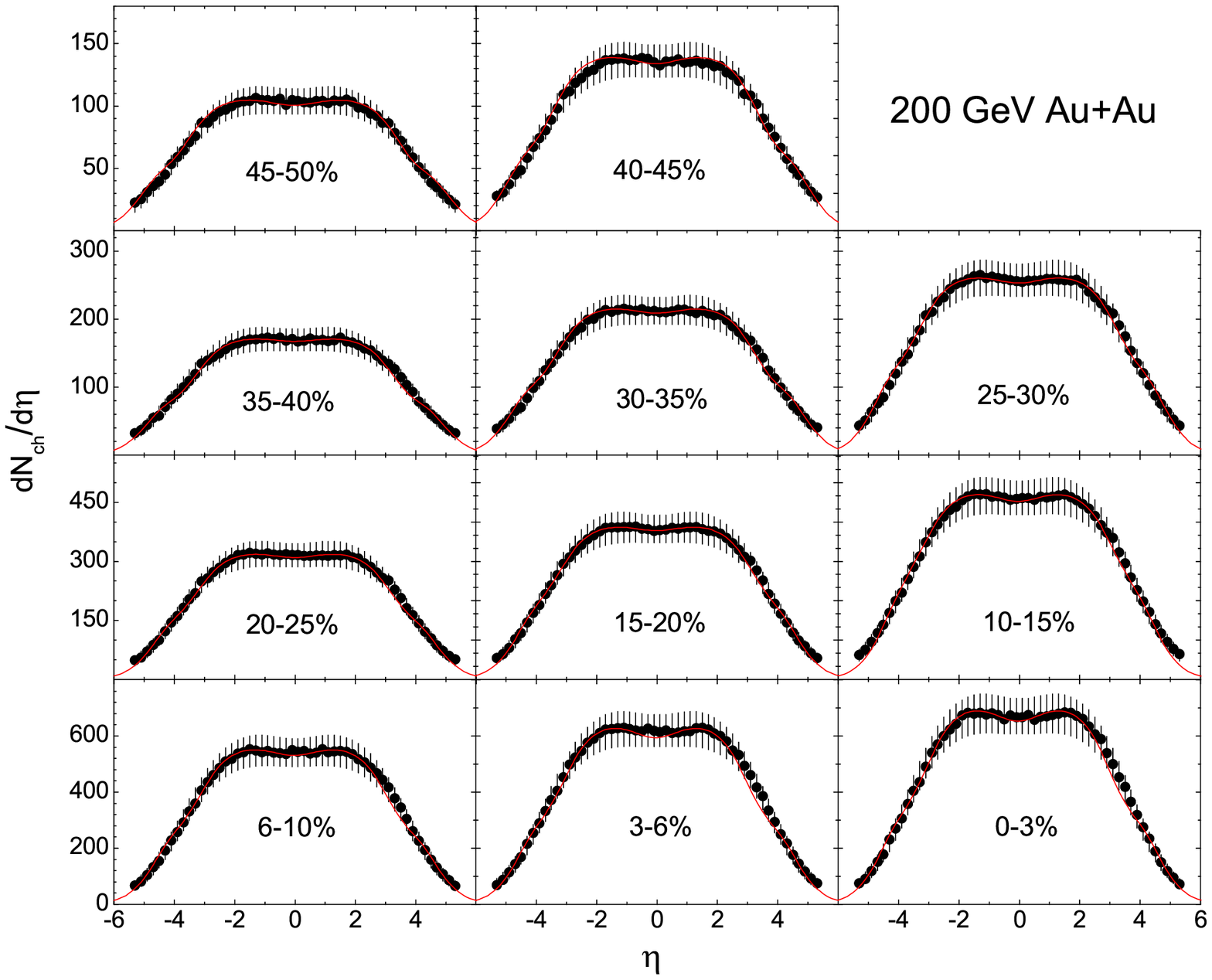}
\end{center} \vskip-1.cm
\caption{(Color online) The charged particle multiplicity
$dN_{ch}/d\eta$ for different centrality bins in Au+Au collisions at
$\sqrt{\mathrm{\it s_{NN}}}$ = 200 GeV. Experimental data of the
PHOBOS Collaboration~\cite{Alver:2010ck} are shown by the scattered
symbols. Our calculated results are shown by the curves.}
\end{figure}

\begin{figure}[htpb]
\begin{center}
\includegraphics[width=0.85\textwidth]{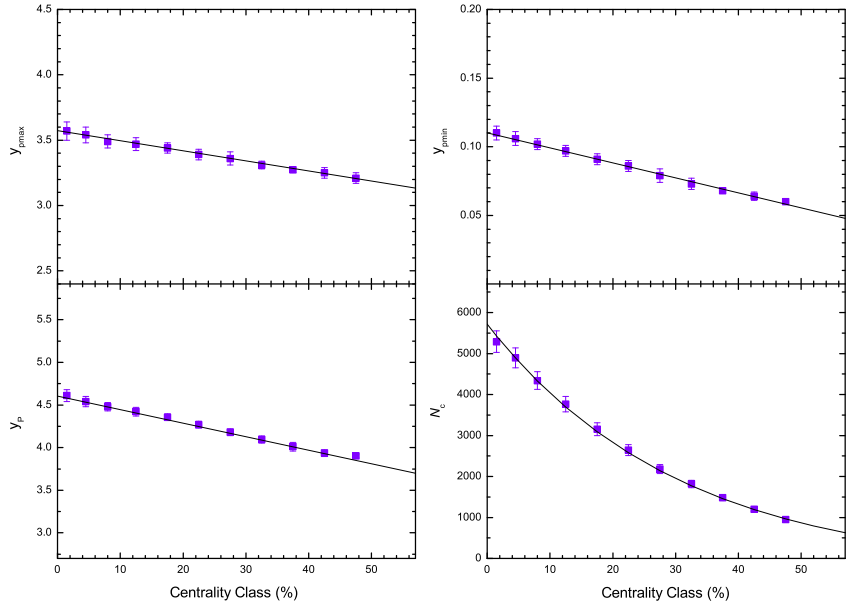}
\end{center}
\vskip-2.cm \caption{(Color online) The dependence of the parameters
on the centrality class. The symbols represent the values used in
the calculations of Fig. 13. The lines are fitted results.}
\end{figure}

\begin{figure}[htbp]
\begin{center}
\includegraphics[width=0.85\textwidth]{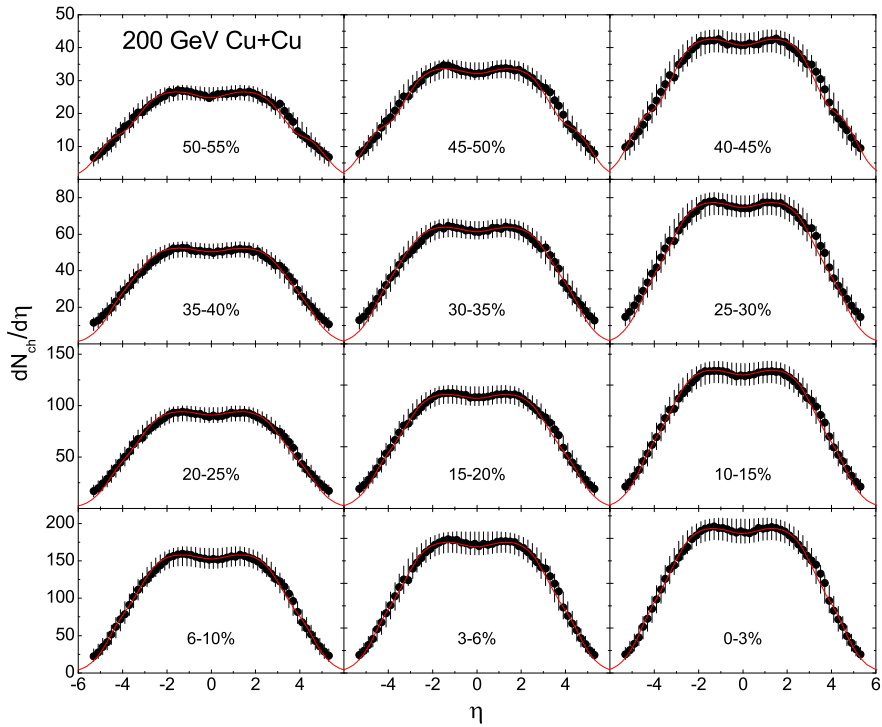}
\end{center}
\vskip-1.1cm \caption{(Color online) Same as for Fig. 13, but for
Cu+Cu collisions.}
\end{figure}

\begin{figure}[bhtp]
\begin{center}
\includegraphics[width=0.8\textwidth]{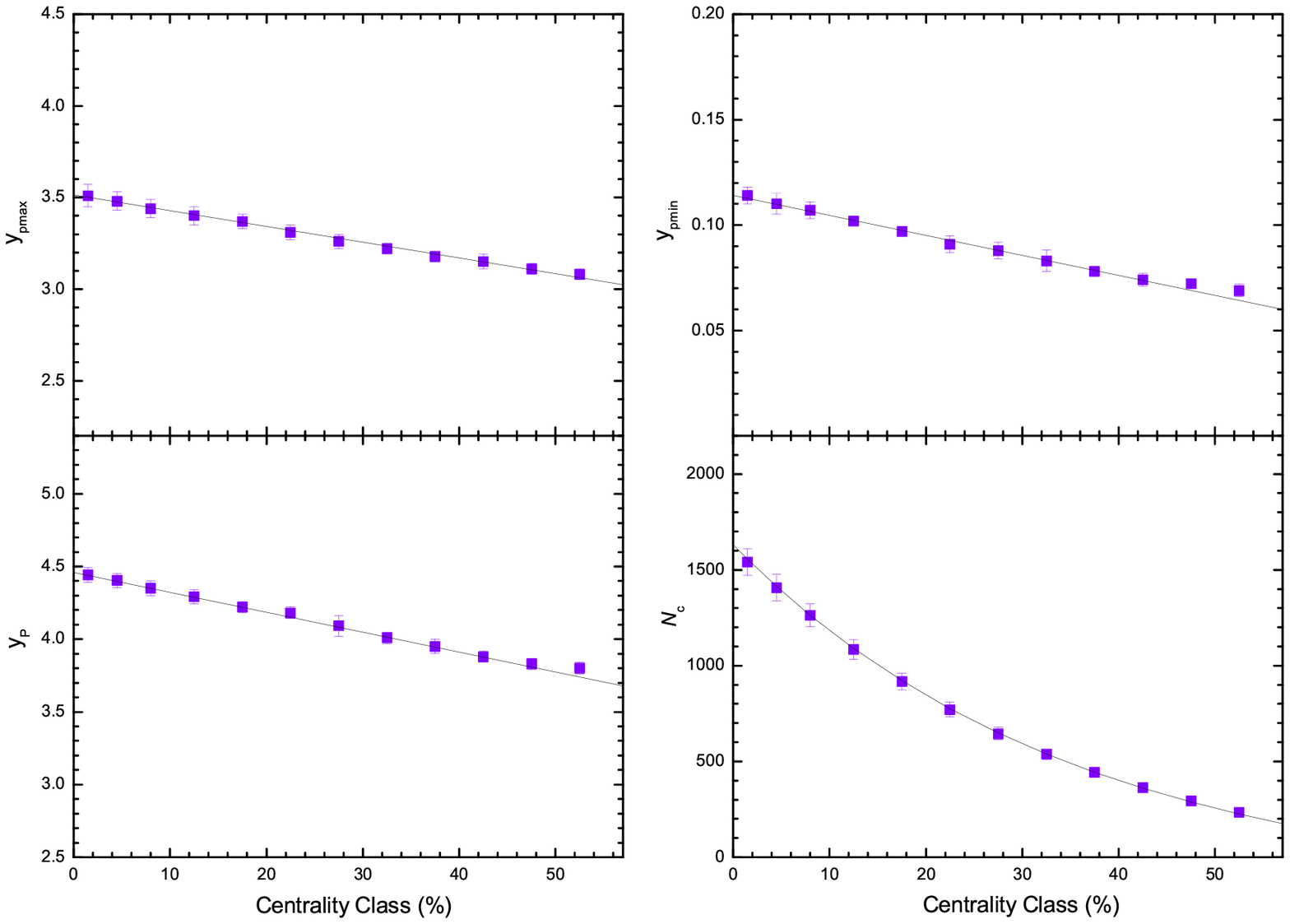}
\end{center}
\vskip-2.cm \caption{(Color online) Same as Fig. 14, but for the
parameters used in  Fig. 15.}
\end{figure}

\begin{figure}[htbp]
\begin{center}
\includegraphics[width=0.8\textwidth]{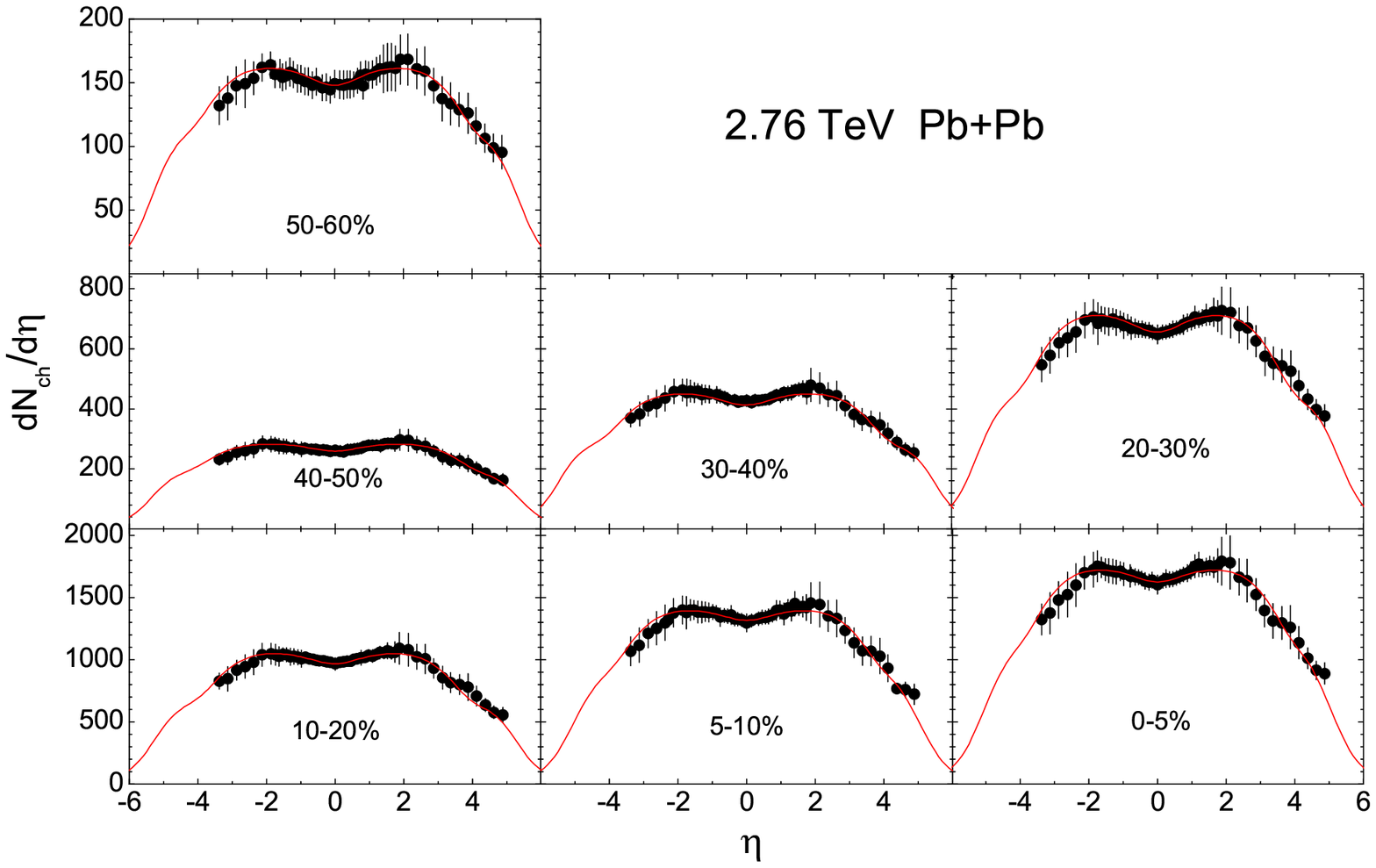}
\end{center}
\vskip-1.cm \caption{(Color online)The charged particle multiplicity
$dN_{ch}/d\eta$ for different centrality bins in Pb+Pb collisions at
$\sqrt{\mathrm{\it s_{NN}}}$ = 2.76 TeV. Experimental data of the
ALICE Collaboration~\cite{Abbas:2013bpa} are shown by the scattered
symbols. Our calculated results are shown by the curves.}
\end{figure}

\begin{figure}[htbp]
\begin{center}
\includegraphics[width=0.85\textwidth]{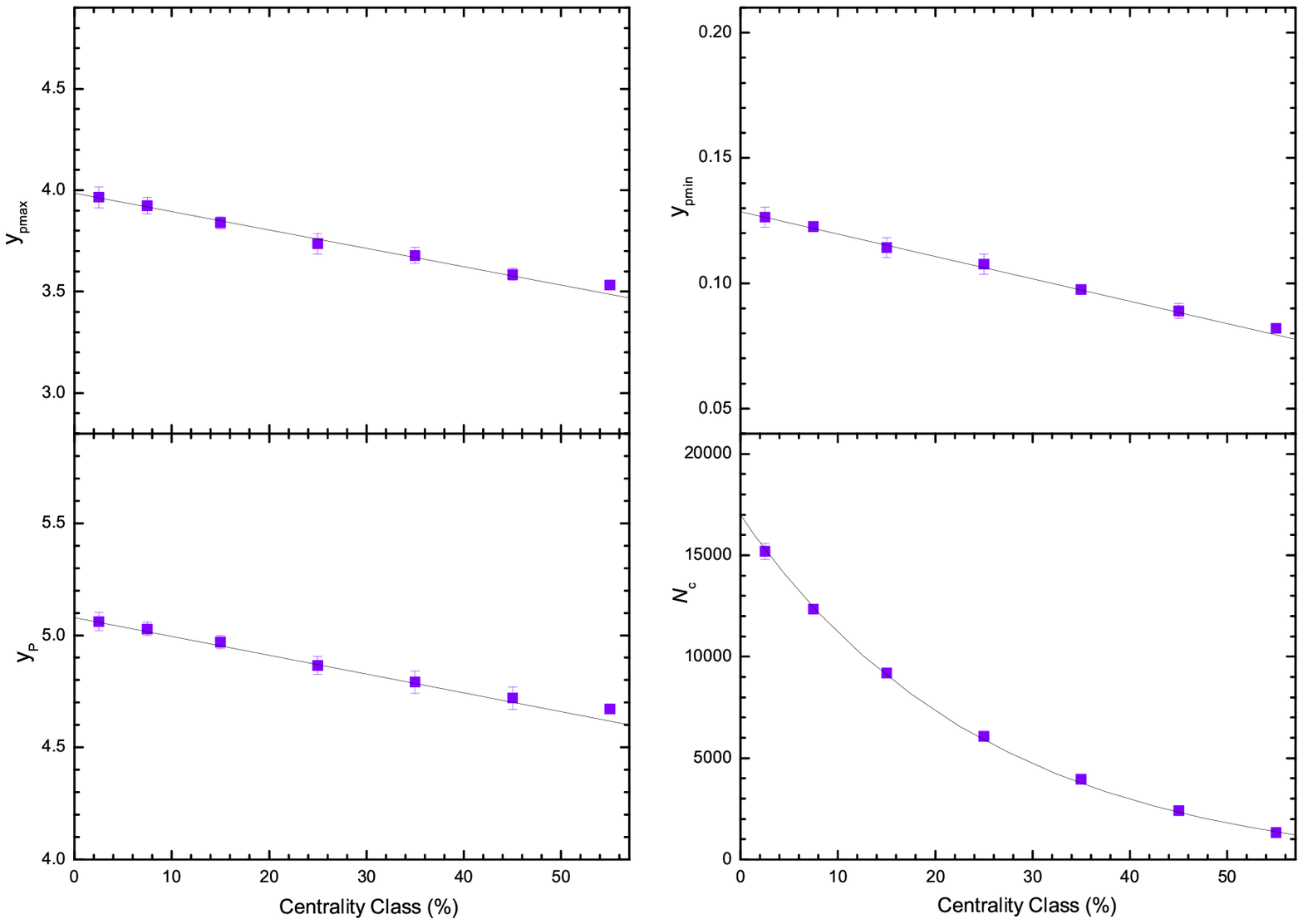}
\end{center}
\vskip-2.cm \caption{(Color online) Same as Fig. 14, but for the
parameters used in Fig. 17.}
\end{figure}


\vskip1.0cm


\begin{thebibliography}{99}
\bibitem{Adams:2003xp}
  J.~Adams {\it et al.}  [STAR Collaboration],
   Phys.\ Rev.\ Lett.\  {\bf 92}, 112301 (2004).
\bibitem{CMS:2008qya}
  The CMS Collaboration,
   CMS-PAS-QCD-08-004.
\bibitem{Abelev:2013haa}
  B.~B.~Abelev {\it et al.}  [ALICE Collaboration],
   arXiv:1307.6796 [nucl-ex].
\bibitem{t1} P.~Braun-Munzinger {\it et al.}, Phys.\ Lett.\ B {\bf 518}, 41 (2001).
\bibitem{t2}J.~Rafelski and J.~Letessier, Nucl.\ Phys.\ A {\bf 715}, 98 (2003).
\bibitem{Andronic:2008gu}
  A.~Andronic, P.~Braun-Munzinger and J.~Stachel,
   Phys.\ Lett.\ B {\bf 673}, 142 (2009)  [Erratum-ibid.\ B {\bf 678}, 516 (2009)].
\bibitem{Cleymans:2006xj}
  J.~Cleymans, I.~Kraus, H.~Oeschler, K.~Redlich and S.~Wheaton,
   Phys.\ Rev.\ C {\bf 74}, 034903 (2006).
\bibitem{BraunMunzinger:2003zz}
  P.~Braun-Munzinger, J.~Stachel and C.~Wetterich,
   Phys.\ Lett.\ B {\bf 596}, 61 (2004).
\bibitem{Adare:2011vy}
  A.~Adare {\it et al.}  [PHENIX Collaboration],
   Phys.\ Rev.\ C {\bf 83}, 064903 (2011)  [arXiv:1102.0753 [nucl-ex]].
\bibitem{Adare:2010fe}
  A.~Adare {\it et al.}  [PHENIX Collaboration],
Phys.\ Rev.\ D {\bf 83}, 052004 (2011).
\bibitem{Abelev:2006cs}
  B.~I.~Abelev {\it et al.}  [STAR Collaboration],
   Phys.\ Rev.\ C {\bf 75}, 064901 (2007)  [nucl-ex/0607033].

\bibitem{Aamodt:2010my}
  KAamodt {\it et al.}  [ALICE Collaboration],
   Phys.\ Lett.\ B {\bf 693}, 53 (2010)  [arXiv:1007.0719 [hep-ex]].
\bibitem{Aad:2010ac}
  G.~Aad {\it et al.}  [ATLAS Collaboration],
   New J.\ Phys.\  {\bf 13}, 053033 (2011)  [arXiv:1012.5104 [hep-ex]].
\bibitem{Khachatryan:2010xs}
  V.~Khachatryan {\it et al.}  [CMS Collaboration],
   JHEP {\bf 1002}, 041 (2010)  [arXiv:1002.0621 [hep-ex]].
\bibitem{Khachatryan:2010us}
  V.~Khachatryan {\it et al.}  [CMS Collaboration],
   Phys.\ Rev.\ Lett.\  {\bf 105}, 022002 (2010).

\bibitem{Cleymans:2013rfq}
  J.~Cleymans, G.~I.~Lykasov, A.~S.~Parvan, A.~S.~Sorin, O.~V.~Teryaev and D.~Worku,
   Phys.\ Lett.\ B {\bf 723}, 351 (2013)  [arXiv:1302.1970
   [hep-ph]]; J.~Cleymans, 
   J.\ Phys.\ Conf.\ Ser.\  {\bf 455}, 012049 (2013);   M.~D.~Azmi and J.~Cleymans,
   arXiv:1311.2909 [hep-ph]; M.~D.~Azmi and J.~Cleymans,
   arXiv:1401.4835 [hep-ph].
\bibitem{Wong:2013sca}
  C.~-Y.~Wong and G.~Wilk,
   Phys.\ Rev.\ D {\bf 87}, 114007 (2013)
    [arXiv:1305.2627 [hep-ph]];  C.~-Y.~Wong and G.~Wilk,
   Acta Phys.\ Polon.\ B {\bf 43}, 2047 (2012)
    [arXiv:1210.3661 [hep-ph]];  G.~Wilk and Z.~Wlodarczyk,
   Phys.\ Rev.\ Lett.\  {\bf 84}, 2770 (2000)  [hep-ph/9908459].




\bibitem{liu3} B.~C.~Li, Y.~Y.~Fu, L.~L.Wang, E.~Q.~Wang and
F.~H.~Liu, J.\ Phys.\ G {\bf 39}, 025009 (2012).
\bibitem{Cleymans:2012ya}
  J.~Cleymans and D.~Worku,
   Eur.\ Phys.\ J.\ A {\bf 48}, 160 (2012).
\bibitem{Feng:2000pca}
  S.~-q.~Feng, F.~Liu and L.~-s.~Liu,
   Phys.\ Rev.\ C {\bf 63}, 014901 (2001).
\bibitem{Liu:2004rm}
  F.~-H.~Liu, N.~N.~Abd Allah and B.~K.~Singh,
   Phys.\ Rev.\ C {\bf 69}, 057601 (2004).


\bibitem{Adare:2008ad}
  A.~Adare {\it et al.}  [PHENIX Collaboration],
   Phys.\ Rev.\ Lett.\  {\bf 101}, 162301 (2008).
\bibitem{Adler:2006wg}
  S.~S.~Adler {\it et al.}  [PHENIX Collaboration],
   Phys.\ Rev.\ Lett.\  {\bf 98}, 172302 (2007).

\bibitem{Alver:2010ck}
  B.~Alver {\it et al.}  [PHOBOS Collaboration],
  Phys.\ Rev.\ C {\bf 83}, 024913 (2011).

\bibitem{Alner:1987wb}
  G.~J.~Alner {\it et al.}  [UA5 Collaboration],
   Phys.\ Rep.\  {\bf 154}, 247 (1987).
\bibitem{Aamodt:2010ft}
  K.~Aamodt {\it et al.}  [ALICE Collaboration],
   Eur.\ Phys.\ J.\ C {\bf 68}, 89 (2010).

\bibitem{Harr:1997sa}
  R.~Harr, C.~Liapis, P.~Karchin, C.~Biino, S.~Erhan, W.~Hofmann, P.~Kreuzer and D.~Lynn {\it et al.},
   Phys.\ Lett.\ B {\bf 401}, 176 (1997).
\bibitem{Abe:1989td}
  F.~Abe {\it et al.}  [CDF Collaboration],
   Phys.\ Rev.\ D {\bf 41}, 2330 (1990).


\bibitem{Back:2003qr}
  B.~B.~Back {\it et al.} [ PHOBOS Collaboration ],
  Phys.\ Lett.\ B {\bf 578}, 297-303 (2004).

\bibitem{Adler:2006hu}
  S.~S.~Adler {\it et al.}  [PHENIX Collaboration],
  Phys.\ Rev.\ Lett.\  {\bf 96}, 202301 (2006).


\bibitem{Abbas:2013bpa}
  E.~Abbas {\it et al.}  [ALICE Collaboration],
   Phys.\ Lett.\ B {\bf 726}, 610 (2013).


\end{thebibliography}
\end{document}